\documentclass{beamer}
\usepackage{graphics,graphicx}
\usepackage{amssymb,amsmath}

\newcommand{\bc}{\begin{center}}
\newcommand{\ec}{\end{center}}
\newcommand{\bd}{\begin{displaymath}}
\newcommand{\ed}{\end{displaymath}}
\newcommand{\be}{\begin{equation}}
\newcommand{\ee}{\end{equation}}
\newcommand{\ba}{\begin{array}}
\newcommand{\ea}{\end{array}}
\newcommand{\bea}{\begin{eqnarray}}
\newcommand{\eea}{\end{eqnarray}}
\newcommand{\bt}{\begin{tabular}}
\newcommand{\et}{\end{tabular}}

\newcommand{\ov}{\overline}
\newcommand{\bp}{\begin{picture}}
\newcommand{\ep}{\end{picture}}
\newcommand{\bfi}{\begin{figure}}
\newcommand{\efi}{\end{figure}}

\begin{document}
\title{New indications of the existence of the\\ 6
top-anti-top quark bound states in\\ LHC experiments}
\author{C.D.~Froggatt, C.R.~Das,\\L.V.~Laperashvili and
H.B.~Nielsen}
\institute{\bf A talk submitted to the Conference ``Quarkonium-2012"\,,\\
Moscow, Russia, RAS-MEPHI, November, 12 - 16, 2012.\\
\alert{\bf Speaker: Larisa Laperashvili}}
\date{14th November, 2012.}
\begin{frame}[plain]
  \titlepage
\end{frame}

\begin{frame}
\frametitle{Abstract}
It was shown: 1) that the mass of the Higgs boson discovered by
LHC corresponds to the stability conditions of the SM vacua and to
the Multiple Point Principle, according to which all vacua of the
SM are degenerate, or almost degenerate; 2) that early predicted
by authors new bound states (NBS) of the 6 top and 6 anti-top
quarks ("T-fireballs"), which are formed by their intermediate
interactions with the Higgs bosons, manifest themselves in the
production of multijets in pp-collisions at LHC. The CMS
experiment of LHC with the production of 10 jets can be explained
by the production of pairs of these NBS along with the production
of pairs of the top-anti-top quarks. 3) Also it was shown that the
next indication of the possible existence of the NBS can be the
decay $H\to \gamma\gamma$, which was observed by CMS-collaboration
of LHC. We have considered the contributions of the one-loop
diagrams to the width $\Gamma_{H\to \gamma\gamma}$, taking into
account the contributions of the T-fireballs along the
contributions of the known SM-particles.
\end{frame}

\begin{frame}
Recent astrophysical measurements give an extremely small value of
the vacuum energy density of the Universe:
$$ {\Large \bf
\rho_{vac}\approx (2.3 \times 10^{-3} \,\, {\rm eV})^4.}
$$
A lot of cosmological theories predict the existence in Nature not
one, but a number (maybe a large number) of vacua.\\[5mm]

According to the Multiple Point Principle (MPP) by
Bennett-Nielsen:\\[5mm]

\alert{\large \bf D.L.~Bennett and H.B.~Nielsen,\\
Int.J.Mod.Phys. {\bf A9} (1994) 5155;}\\[5mm]
all vacua in the Universe are degenerate, or almost degenerate,
e.g. have an extremely small cosmological constant.
\end{frame}

\begin{frame}
In the assumption that there exist two vacua in the Standard Model
(SM), when the effective Higgs potential has the two degenerate
minima:\\ the first one - at the Electroweak scale:  
${\Large \bf \phi_{min1}\approx 246\, \,{\rm GeV}}$,\\
and the second one - at the Planck scale : ${\Large \bf
\phi_{min2}\sim M_{Pl}}$,\\[-15mm]

\centering{\includegraphics[height=80mm,keepaspectratio=true,angle=0]{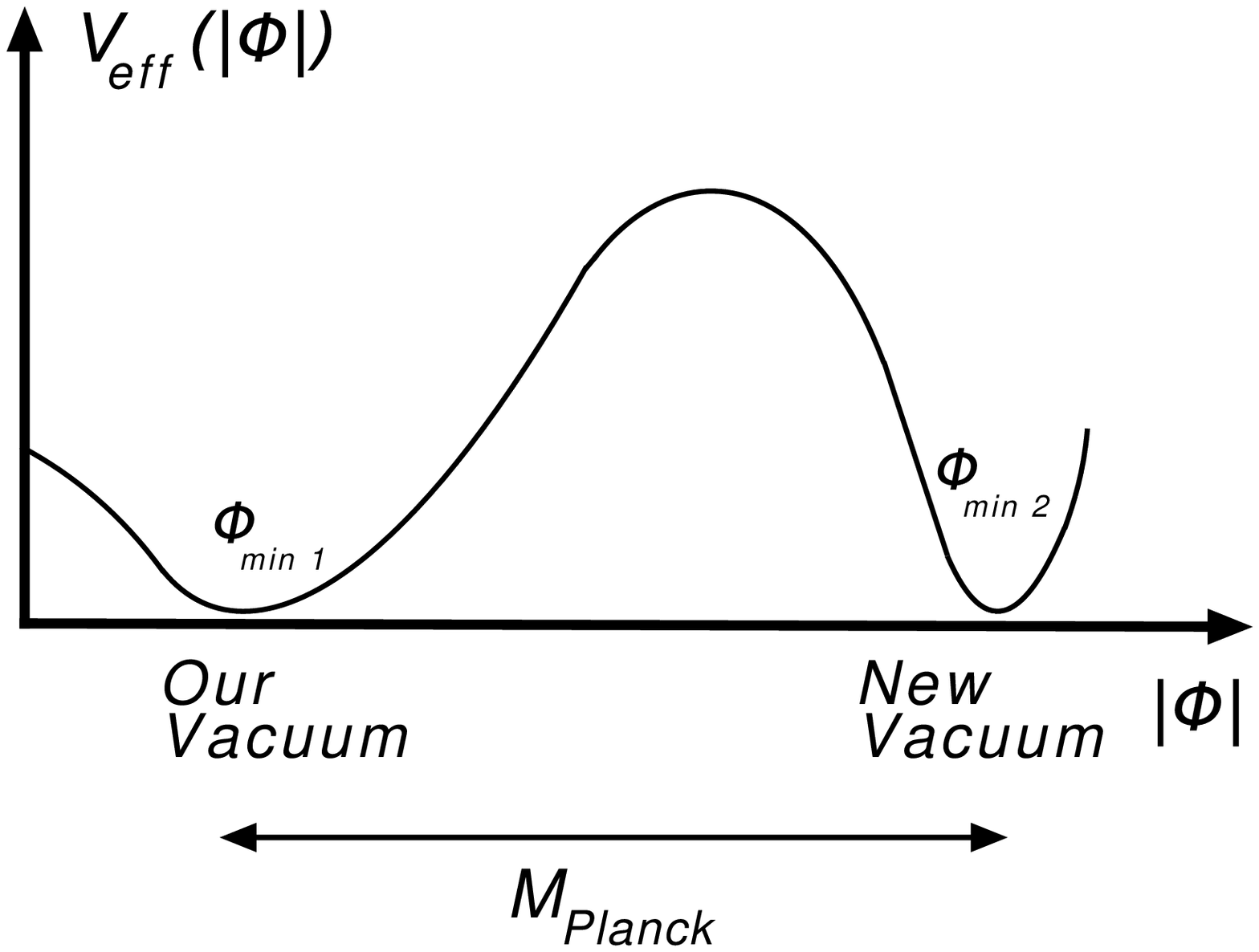}}
\end{frame}

\begin{frame}
Froggatt and Nielsen\\[5mm]

\alert{\large \bf C.D.~Froggatt, H.B.~Nielsen,\\
Phys.Lett. B {\bf 368}, 96 (1996)}\\[5mm]
obtained the following prediction for the masses of the top-quark
and Higgs boson :
$$ {\Large \bf  M_t = 173 \pm 5\,\,{\mbox{GeV}},\quad M_H =
135 \pm 10\,\,{\mbox{GeV}}.} $$ under the stability condition of
the both vacua.\\[5mm]

The prediction of the top-quark mass was confirmed very soon by
experiment.
\end{frame}

\begin{frame}
High a value of the Higgs mass was explained recently by the
radiative corrections to the Higgs effective potential:\\[5mm]
\alert{\large \bf G.~Degrassi, S.~ Di~Vita, J.~Elias-Miro, J.~R.~
Espinosa, G.~F.~Giudice, G.~Isidori, A.~Strumia, JHEP
1208:098,2012; arXiv:1205.6497.}\\[5mm]

Previously we also considered the radiative corrections to
the Higgs effective potential:\\[5mm]
\alert{\large \bf C.D. Froggatt, L.V. Laperashvili, H.B. Nielsen,
Phys.Atom.Nucl. 69 (2006) 67; e-Print: hep-ph/0407102,}\\[5mm]
explicitly shown the existence of the second minimum at the Planck
scale.
\end{frame}

\begin{frame}
The investigation by G.~Degrassi et al. has led to the following
prediction of the Higgs mass:
$${\Large \bf    M_H = 129 \pm 2 \,\,{\mbox{GeV}},}$$
which is close to the value of the Higgs boson mass found by LHC:
$${\Large \bf    M_H \approx 126.4 \,\,{\mbox{GeV}}.}$$
And now when the existence of the Higgs boson is conclusively
proven, the existence of the new type of the top-quark bound
states becomes possible: they are created by the Higgs boson
virtual exchanges between top-quarks ${\Large \bf  tt,\,t\bar
t,\,}$ è ${\Large \bf  \bar t\bar t}$, which ensure the
attraction:
\end{frame}

\begin{frame}
\centering{\includegraphics[height=50mm,keepaspectratio=true,angle=0]{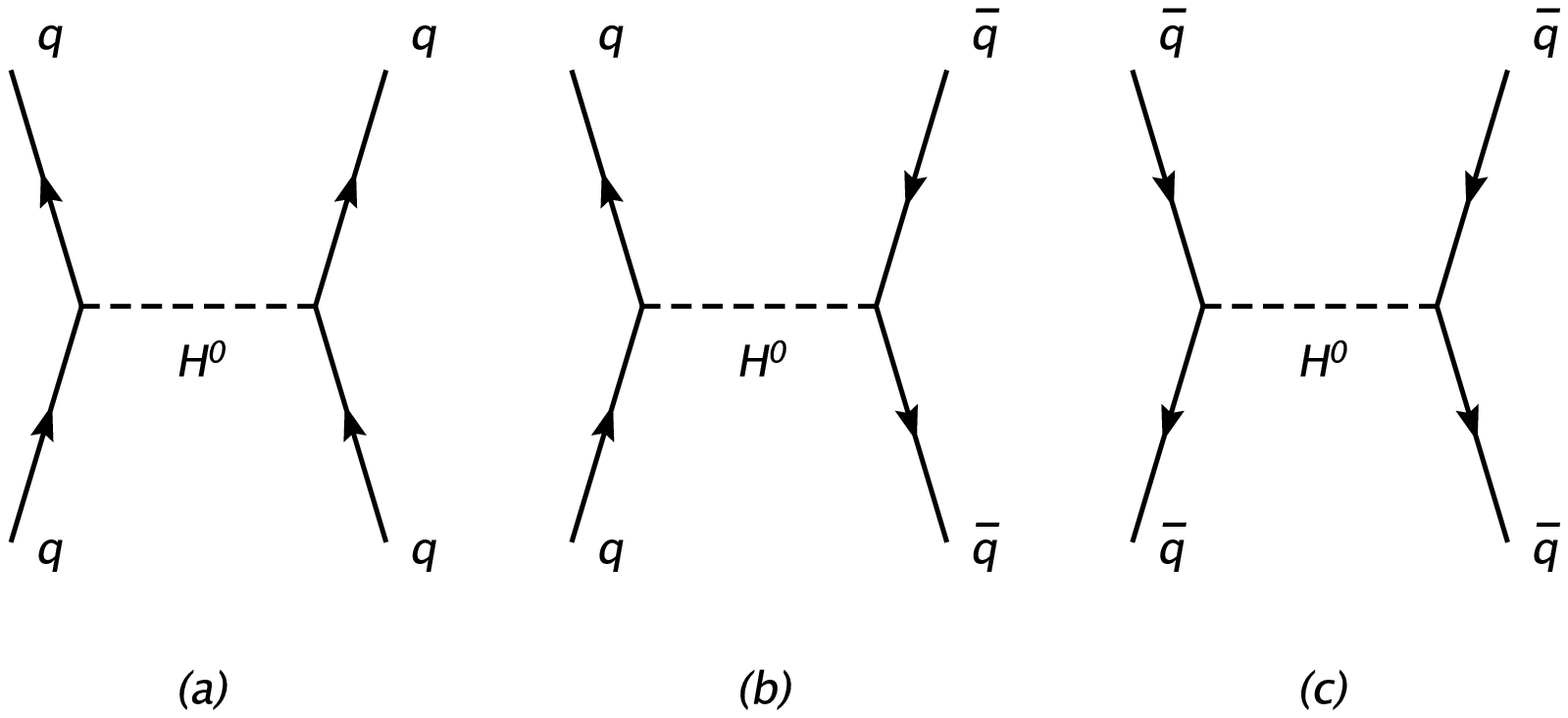}}

This occurs because the Yukawa coupling constant ${\Large \bf
g_t}$ in the Lagrangian, describing the interaction of top-quarks
with the Higgs bosons:
$${\Large \bf
 L = \frac 12 D_{\mu}\Phi_H D^{\mu}\Phi_H + \frac{g_t}{\sqrt
          2}\ov{\psi_{tL}}\psi_{tR}\Phi_H  + h.c.}$$
is large enough: ${\Large g_t\sim 1}$.
\end{frame}

\begin{frame}
In the Reference:\\
\alert{\large \bf C.D.~Froggatt and H.B.~Nielsen. {\it
Invited talk by\\ H.B.~Nielsen at the ``XXXI ITEP Winter School of
Physics''\\ (February 18--26, 2003, Moscow, Russia)};\\ arXiv: hep-ph/0308144.}\\
was first assumed that\\$\bullet$  there exists ${\Large \bf
1S}$-bound state ${\Large \bf
6t+6\bar t}$, which is a scalar particle and color singlet;\\
$\bullet$ that the forces responsible for the forming of these
bound states originate from the virtual exchanges of the Higgs
bosons between
top-quarks;\\
$\bullet$ that these forces are so strong that almost compensate
the mass of
12 top-quarks which are contained in these bound states;\\
$\bullet$ that there exists also a new bound state ${\Large \bf 6t
+ 5\bar t}$, which is a fermion similar to the quark of the 4th
generation having quantum numbers of ${\Large \bf t}$-quark.
\end{frame}

\begin{frame}

We call these bound states "T-fireballs": \bc ${\Large \bf
T_s=6t+6\bar t}$

and

${\Large \bf  T_f=6t + 5\bar t.}$ \ec

The explanation of the stability of the bound state
$${\Large \bf 6t+6\bar t}$$
is given by the Pauli principle: top-quark has two spin and three
color degrees of freedom (total 6 DOF). By this reason, 6 quarks
have the maximal binding energy, and 6 pairs of ${\Large \bf t\bar
t}$ in ${\Large \bf 1S}$-wave state create a stable colorless
scalar bound state ${\Large \bf T_s}$. It is obvious that ${\Large
\bf 2S}$-state corresponds to the forming of the more heavy bound
state of the 6 top-anti-top quarks, etc.
\end{frame}

\begin{frame}
The next investigations of the properties of new bound states of
the 6 top-anti-top-quarks (calculations of their mass values,
experimental searching, etc.,) were made by the following authors:\\[5mm]
{\large \bf C.R.~Das, C.D.~Froggatt, L.V.~Laperashvili,
R.B.~Nevzorov and H.B.~Nielsen:}

\begin{itemize}
\item \alert{\large \bf C.D.~Froggatt, H.B.~Nielsen, Phys.Rev. D80
(2009) 034033: arXiv:0811.2089;}\\

\item \alert{\large \bf C.D.~Froggatt, H.B.~Nielsen, Presented at
Conference: C08-07-30, arXiv:0810.0475 [hep-ph];}\\

\item \alert{\large \bf  C.D.~Froggatt, H.B.~Nielsen, L.V.
~Laperashvili, Int.J.Mod.Phys. A20 (2005) 1268; e-Print:
hep-ph/0406110;}\\

\item \alert{\large \bf  C.D.~Froggatt, L.V.~Laperashvili, H.B.~
Nielsen. Talk given at Conference: C04-05-24; e-Print:
hep-ph/0410243.}
\end{itemize}
\end{frame}

\begin{frame}
\begin{itemize}
\item \alert{\large \bf  C.D.~Froggatt, L.V.~Laperashvili,  R.B.~
Nevzorov, H.B.~Nielsen, C.R.~Das, CHEP-PKU-1-04-2008;
arXiv:0804.4506;} \\

\item \alert{\large \bf C.R.~Das, C.D.~Froggatt, L.V.~Laperashvili,
H.B.~Nielsen, Proceedings of the Fourteenth Lomonosov Conference
on Elementary Particle Physics, 379 - 381. Editor: Alexander I.
Studenikin, (World Scientific Publishing
Company, 2010); arXiv:0908.4514 [hep-ph];}\\

\item \alert{\large \bf  C.R.~Das, C.D.~Froggatt, L.V.~ Laperashvili,
H.B.~Nielsen, Int.J.Mod.Phys. A26 (2011) 2503;
arXiv:0812.0828 [hep-ph].}\\
\end{itemize}
\end{frame}

\begin{frame}
The estimate of the mass ${\Large \bf M_s}$ and ${\Large \bf M_f}$
of the fireballs ${\Large \bf T_s=6t+6\bar t}$ and ${\Large \bf
T_f=6t + 5\bar t}$ gave the following result: \bc ${\Large \bf 63
\,\, GeV < M_s \lesssim 300\,\, GeV, \qquad \large \bf M_f \approx
600-700\,\,GeV.}$  \ec If the mass ${\Large \bf M_s}$ would be
less than a half of the Higgs mass  (${\Large \bf M_s < M_H/2}$),
then in the LHC experiments we could observe the decays of the
Higgs boson into the two scalar fireballs:
$${
\Large \bf  H \to 2T_s. }$$ However, we don't see such decays.

Instead, we observe in the CMS-collaboration experiments of LHC
the production of 10 jets:
\end{frame}

\begin{frame}
\centering\includegraphics[height=90mm,keepaspectratio=true,angle=0]{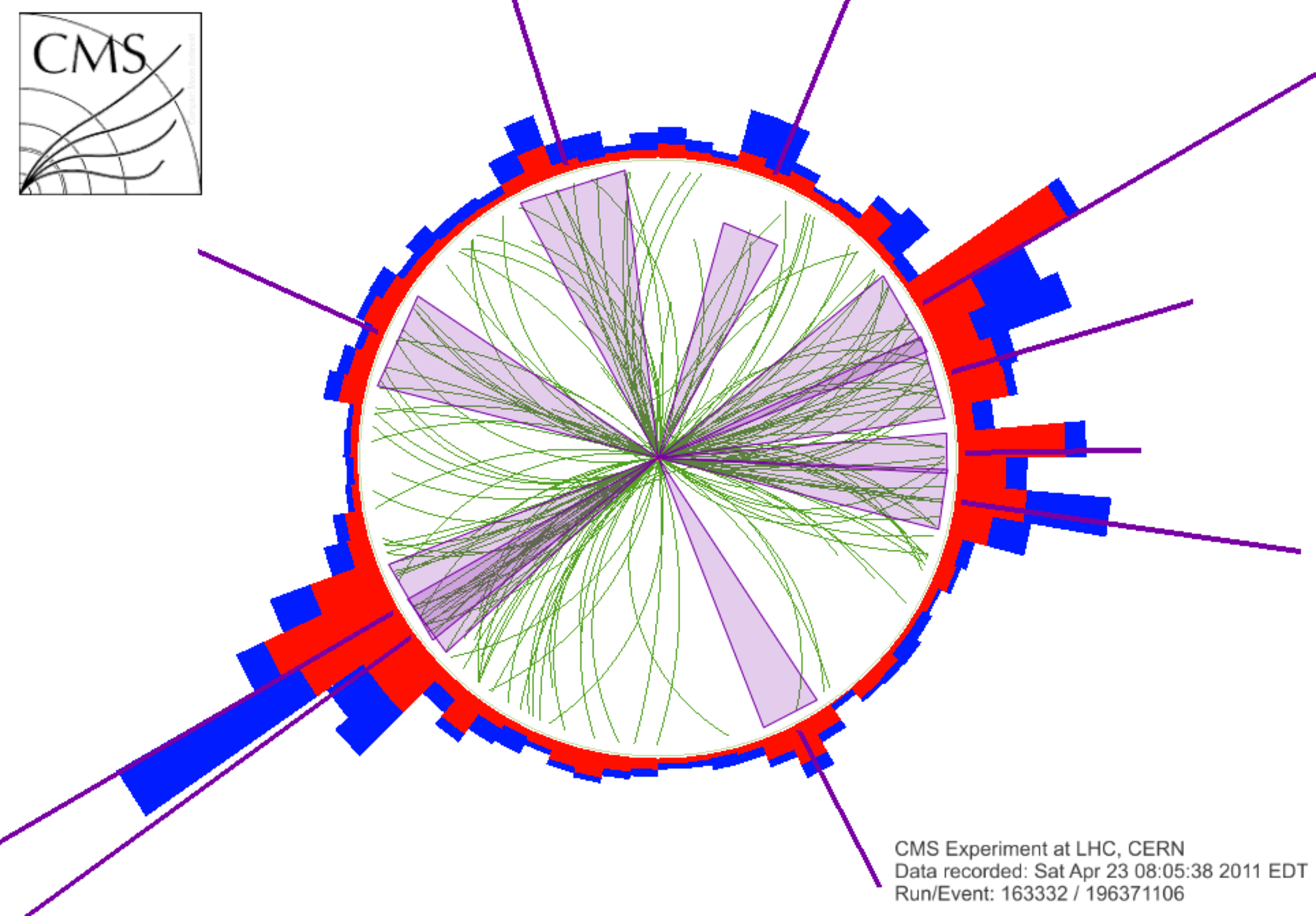}
\end{frame}

\begin{frame}
The production of 10 jets can be easy explained by the
participation of T-fireballs in the two-gluon interaction with
protons in collisions of protons in the LHC experiments.\\[5mm]

\hspace{0.7cm}\includegraphics[height=70mm,keepaspectratio=true,angle=0]{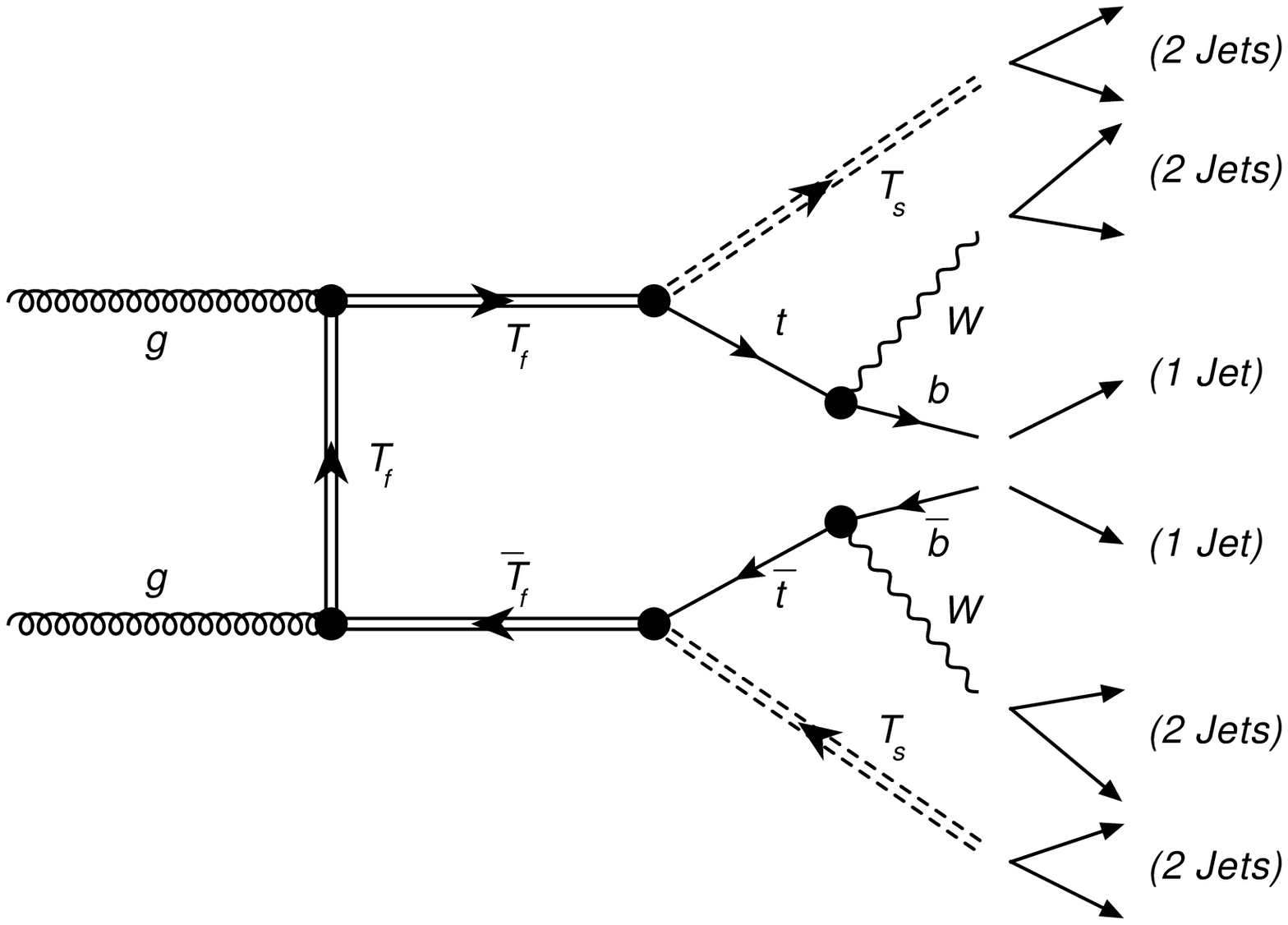}
\end{frame}

\begin{frame}
The top-quark, shown in the picture, decays by the following way:
$${\Large \bf t\to W + b.}$$
As a result, in the up side of the diagram we have the production
of the 5 jets according to the scheme:
\bc

 ${\Large \bf W \to q\bar q \to}$ 2 jets,

 ${\Large \bf b \to}$ 1 jet,

${\Large \bf T_s \to q\bar q \to}$ 2 jets \\(presumably, the
process is going via ${\Large \bf T_s \to b\bar b}$).

\ec

The same number of jets is created in the down side of the diagram.\\[5mm]
Thus, the experimental observation of the 10 jets can be explained
by the production of T-fireballs in the proton collisions at LHC.
\end{frame}

\begin{frame}
LHC experiment established the dependence of the events
(${\Large \bf n}$) of the production of the ${\Large \bf N}$ jets:\\
\alert{\large\bf Nick van Remortel, {\it Past, Present and Future
of QCD}, CERN, 2012.}

This dependence is presented by the following histogram:\\
\hspace{1.8cm}\includegraphics[height=100mm,keepaspectratio=true,angle=0]{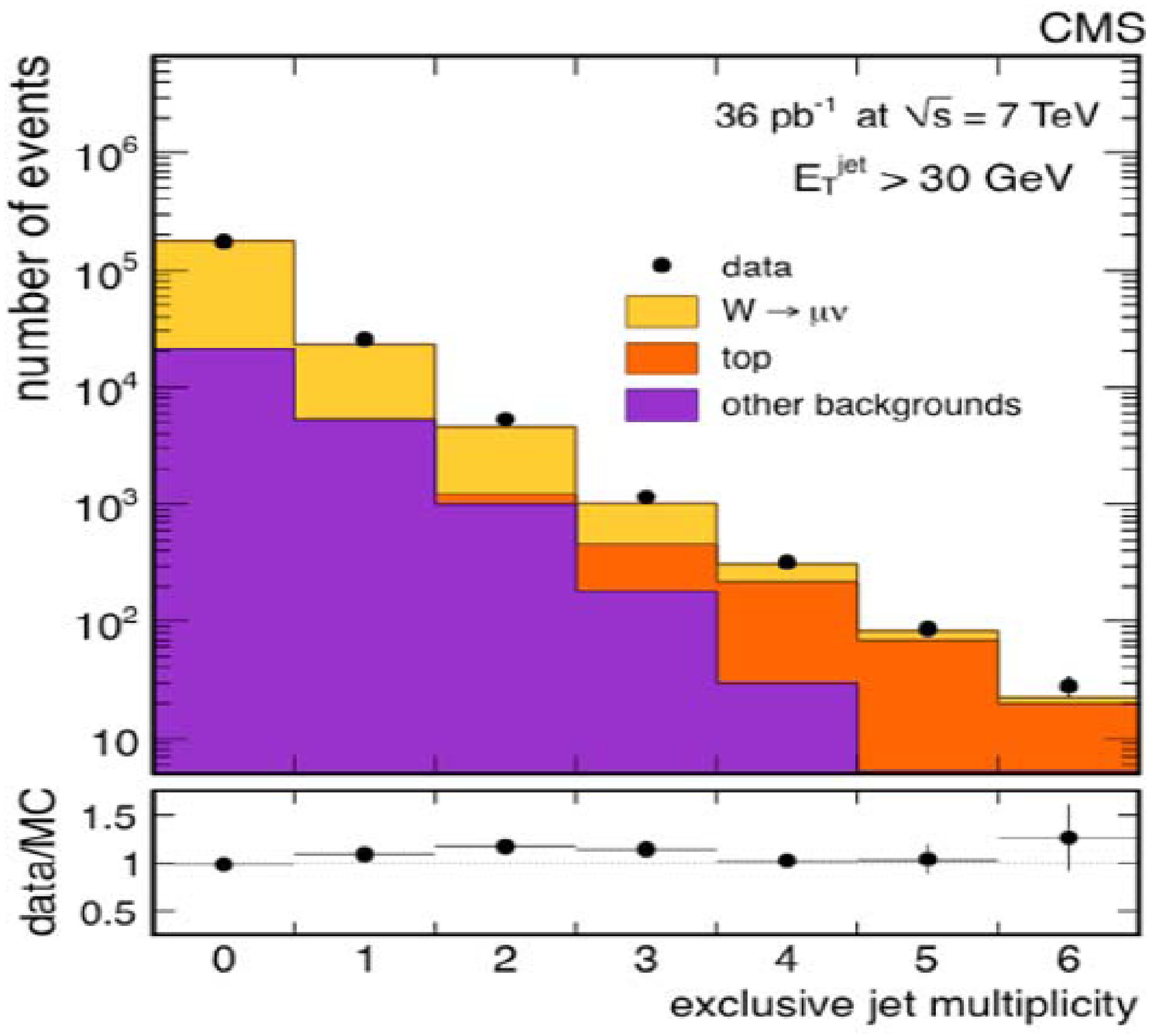}
\end{frame}

\begin{frame}
Using this experimentally established histogram, we determined the
dependence of the probability ${\Large \bf P(N)}$
of the production of jets vs their number ${\Large \bf N}$.\\[5mm]
Such a dependence is given by the following curve:
$$ {\Large \bf \log P(N)\approx -0.05 - 0.80 N + 0.01 N^2. }$$
Below figure presents the dependence ${\Large \bf W = |\log
P(N)|}$\,vs\,${\Large \bf N}$:

\hspace{1.8cm}
\includegraphics[height=50mm,keepaspectratio=true,angle=0]{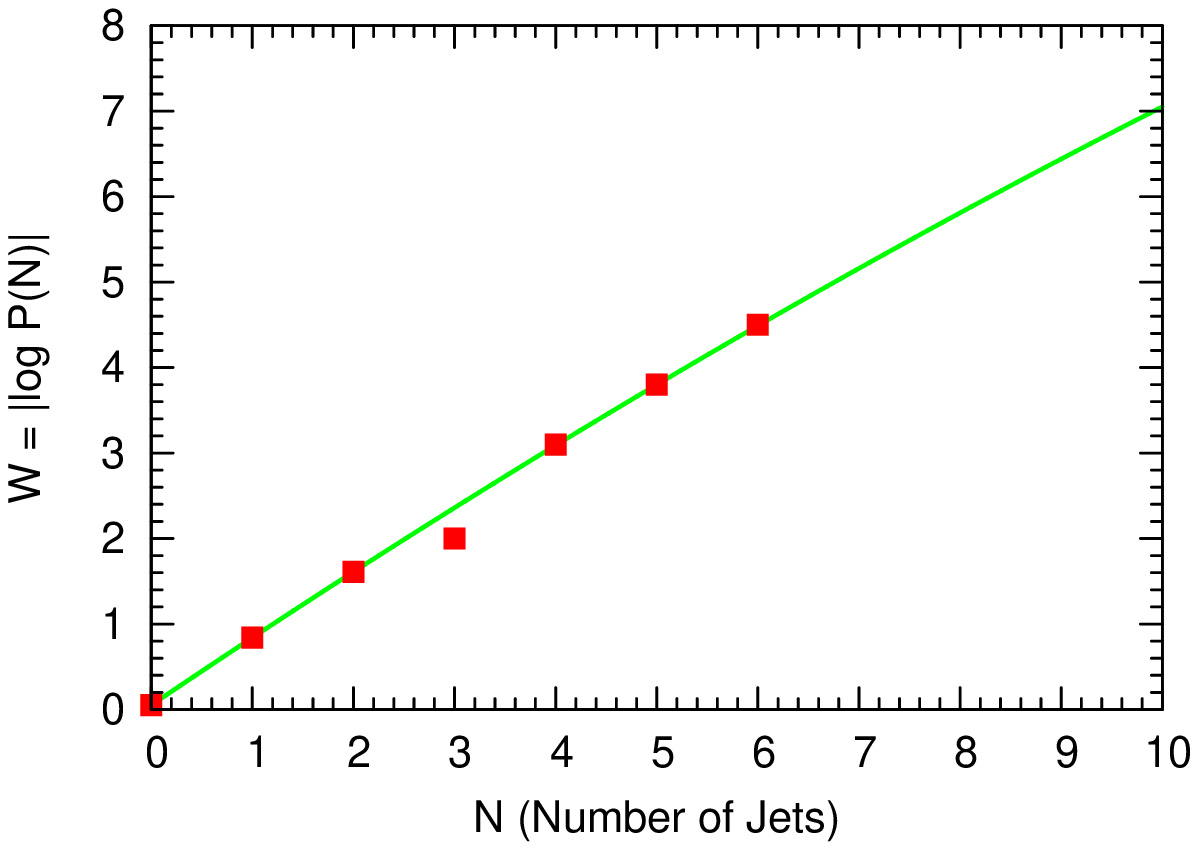}
\end{frame}

\begin{frame}
Extrapolating this curve to the point ${\Large \bf N = 10}$ we
obtained the probability of the production of 10 jets.\\[5mm]
It is equal to:
$${\Large \bf
     P(10) \sim 10^{-7},}$$
e.g. extremely small: one 10-jets event on the 10 millions of the
total events.

But the existence of the T-fireballs essentially increases the
probability of the observation of the 10-jets process. Maybe, just
this is a reason why the 10-jets event is seen at LHC.
\end{frame}

\begin{frame}
The next indication of the possible existence of the new bound
states of 6 top-anti-top quarks can be the decay of the Higgs
boson into the two gamma-quants, which was observed by
CMS-collaboration of LHC:
$${\Large \bf    H \to \gamma\gamma.}$$
The result of the experimental measurements showed that the ratio
of this experimental width ${\Large \bf \Gamma_{exp}}$ to the same
width calculated theoretically in the SM, is not equal to unity,
but is given by the following estimate:
$$\Large \bf \frac{\Gamma_{exp}}{\Gamma_{SM}} = 1.56\pm 0.43$$ -
by CMS, and
$$\Large \bf \frac{\Gamma_{exp}}{ \Gamma_{SM}} = 1.8\pm 0.5$$ -
by ATLAS.
\end{frame}

\begin{frame}
We have considered the contributions of the one-loop diagrams to
the width of ${\Large \bf \Gamma_{H\to \gamma\gamma}}$, taking
into account the contributions of T-fireballs ${\Large \bf T_s}$
and ${\Large \bf T_f}$ along the contributions of the SM-particles
${\Large \bf W,\,H}$ and top-quark ${\Large \bf t}$:\\[1.7cm]
\centering
\includegraphics[height=31mm,keepaspectratio=true,angle=0]{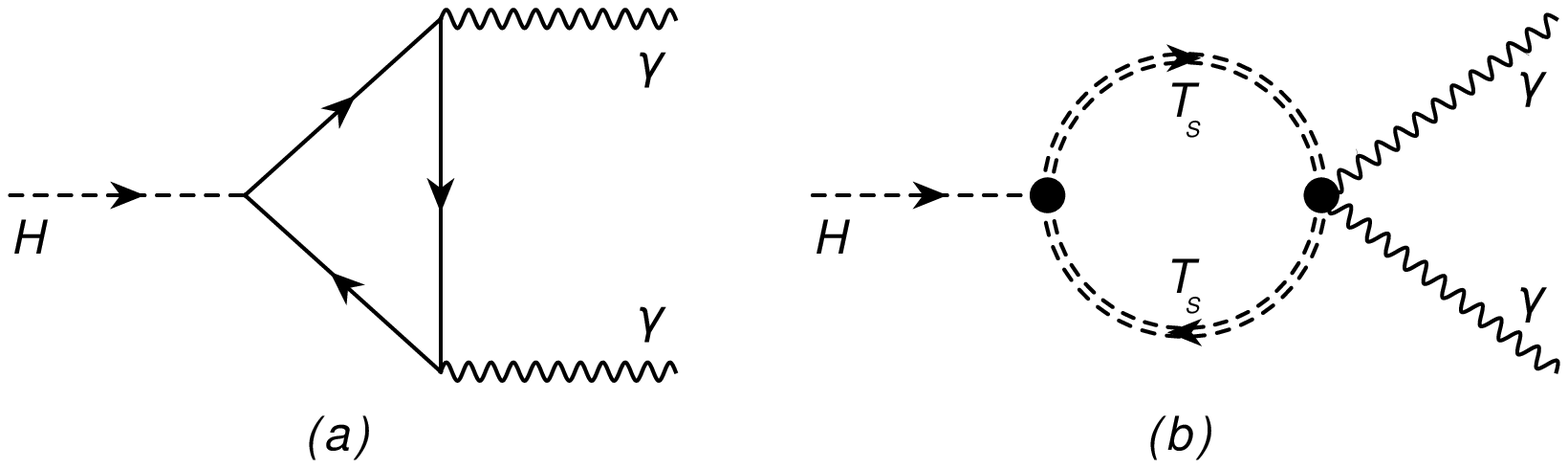}
\end{frame}

\begin{frame}
The calculations give the following result:\\[5mm]
$${\Large \bf \frac{\Gamma_{exp}}{ \Gamma_{SM}} \approx
\left|\frac{A_{W,H} + A_t + A_{T_f} +A_{T_s}}{A_{W,H} +
A_t}\right|^2,}$$
 where ${\Large \bf A_X}$ are the amplitudes of
different contributions. We have chosen:
$${\Large \bf  A_{W,H} + A_t = 1.}$$
 This is the SM contribution.
Calculations give:\\
$${\Large \bf A_{W,H}\approx + 1.14, \quad A_t\approx - 0.14,}$$
$${\Large \bf  A_{T_f}\approx - 0.89, \quad  A_{T_s}\approx +
1.10.}$$
\end{frame}

\begin{frame}
The total result is:\\
$${\Large \bf \frac{\Gamma_{exp}}{ \Gamma_{SM}} \approx
(1.21)^2\approx 1.46.}$$ We have used in calculations the
following values of masses: ${\Large \bf M_s = 300}$ GeV and
${\Large
\bf M_f = 700}$ GeV.\\[5mm]
Although the effect of the T-fireballs is obvious, the result
depends on the masses of T-fireballs, which are unknown.\\
The last word is left for the LHC experiments.
\end{frame}

\end{document}